\documentclass[preprint]{aastex}

\begin{document}

\title {The Eclipsing System EP Andromedae and its Circumbinary Companions}
\author{Jae Woo Lee, Tobias Cornelius Hinse, and Jang-Ho Park}
\affil{Korea Astronomy and Space Science Institute, Daejeon 305-348, Korea}
\email{jwlee@kasi.re.kr, tchinse@gmail.com, pooh107162@kasi.re.kr}

\begin{abstract}
We present new long-term CCD photometry for EP And acquired during the period 2007 to 2012. The light curves display total eclipses
at primary minima and season-to-season light variability. Our synthesis for all available light curves indicates that the eclipsing pair 
is a W-type overcontact binary with parameters of $q$=2.578, $i$=83$^\circ$.3, $\Delta T$=27 K, $f$=28 \%, and $l_3$=2$\sim$3 \%. 
The asymmetric light curves in 2007 were satisfactorily modeled by a cool spot on either of the eclipsing components from 
a magnetic dynamo. Including our 95 timing measurements, a total of 414 times of minimum light spanning about 82 yr were used for 
a period study. A detailed analysis of the eclipse timing diagram revealed that the orbital period of EP And has varied as a combination 
of an upward-opening parabola and two periodic variations, with cycle lengths of $P_3$=44.6 yr and $P_4$=1.834 yr and semi-amplitudes 
of $K_3$=0.0100 d and $K_4$=0.0039 d, respectively. The observed period increase at a fractional rate of $+$1.39$\times$10$^{-10}$ 
is in excellent agreement with that calculated from the W-D code and can be plausibly explained by some combination of mass transfer 
from the primary to the secondary star and angular momentum loss due to magnetic braking. The most reasonable explanation for both cycles 
is a pair of light-travel-time effects driven by the possible existence of a third and fourth component with projected masses of 
$M_3$=0.25 M$_\odot$ and $M_4$=0.90 M$_\odot$. The more massive companion could be revealed using high-resolution spectroscopic data 
extending over the course of a few years and could also be a binary itself. It is possible that the circumbinary objects may have 
played an important role in the formation and evolution of the eclipsing pair, which would cause it to have a short initial orbital 
period and thus evolve into an overcontact configuration by angular momentum loss.
\end{abstract}

{}

\section{INTRODUCTION}

Many overcontact binaries are thought to have evolved from initially detached binaries by angular momentum loss (AML) via magnetic torques 
from stellar winds and to end with coalescence of both components into single stars (Bradstreet \& Guinan 1994; Pribulla \& Rucinski 2006). 
In this scenario, the orbital angular momentum is tidally coupled to the spin angular momentum. In order for the spin-orbit coupling 
to work efficiently, the initial orbital periods should be shorter than about 5 days. Circumbinary objects in multiple systems may remove 
angular momentum from the central pairs via Kozai oscillation (Kozai 1962; Pribulla \& Rucinski 2006) or a combination of the Kozai cycle 
and tidal friction (Fabrycky \& Tremaine 2007) and form the initial tidal-locked detached binaries with a short orbital period. 

The statistical study of Pribulla \& Rucinski (2006) suggests that a large percentage of close binares reside in triple or multiple systems. 
The presence of a third body orbiting an eclipsing close binary causes a periodic variation of the eclipsing period due to the increasing 
and decreasing light-travel times (LTT) to the observer (Irwin 1952, 1959). The LTT effect can be studied by a detailed analysis of
eclipse timing diagram, showing the differences between the observed ($O$) and the calculated ($C$) timings of minimum light {\it versus}
time (e.g., Lee et al. 2009a, 2013). In order to advance this subject, we have been observing short-period eclipsing binaries, 
such as overcontact and near-contact systems. In this work, we choose the W UMa-type binary EP And, because the list of eclipse timings 
goes back far enough to understand the binary's period behavior. 

Since the discovery of its variability by Strohmeier et al. (1955), EP And (TYC 2827-17-1, GSC 2827-17, 2MASS J01422933+4445424; 
$V$=+11.90, $B$--$V$=+0.57) has been the topic of several investigations, but its properties are relatively poorly known compared to
other short-period binaries. Most recently, Manzoori (2012) reviewed the observational history of the system and presented the first 
comprehensive photometric study. The author analyzed the photometric data from both the AAVSO (American Association of Variable Stars Observers) 
International Database\footnote {http://www.aavso.org/} and the WASP (Wide Angle Search for Planets) public archive (Butters et al. 2010), 
and concluded that EP And is an A-type (defined observationally by Binnendijk 1970) overcontact binary with a mass ratio of $q$=0.395, 
an orbital inclination of $i$=81$^\circ$.5, a temperature difference between the components of $\Delta T$=120 K, and a fill-out factor 
of $f$ = 16.6 \%. Here, $q$ is the ratio of the secondary'mass to that of the primay star (eclipsed at the primary minimum) and 
$f$=($\Omega_{\rm in}$--$\Omega$)/($\Omega_{\rm in}$--$\Omega_{\rm out}$), where the potentials $\Omega_{\rm in}$ and 
$\Omega_{\rm out}$ define the inner and outer critical surfaces in Roche geometry and $\Omega$ is the potential corresponding 
to the surface of the overcontact binary. From the analysis of eclipse timings including five epochs determined from the AAVSO data, 
he suggested that the orbital period of EP And can be sorted into a secular period decrease caused by mass transfer between 
the components, an LTT effect due to the orbit of a third body with a period of 41.2 yr and a minimum mass of 0.15 M$_\odot$, and 
the effect of magnetic activity with a cycle length of 11.7 yr. 

In order to obtain a unique set of photometric solutions and to examine whether the Manzoori's (2012) suggestion is appropriate 
for the orbital period change, we have studied in detail the long-term photometric behavior of EP And from all available data. 
Our results from both the light-curve synthesis and the orbital period study show that EP And is probably a multiple system.
This study follows the following structure. In section 2, we present our new photometric observations of EP And. 
Absolute dimensions of the eclipsing pair are determined from light-curve modeling in section 3. In section 4, we present 
an orbital period study of the eclipse timing diagram and determine two significant periods in the timing measurements. 
In the last section we give a discussion highlighting that the two periodic variations are most likely attributed to the presence 
of circumbinary companions.

\section{NEW LONG-TERM CCD PHOTOMETRY}

New CCD photometric observations of EP And were obtained between 2007 and 2012, using CCD cameras and a $BVR$ filter set attached 
to the 61-cm reflector at Sobaeksan Optical Astronomy Observatory (SOAO) in Korea. The observations of the first two seasons were 
carried out to secure complete multiband light curves and the others to collect additional eclipse timings. A summary of 
the observations is listed in Table 1, where we present observing interval, numbers of nights, CCD type, and field of view (FOV). 
The instruments and reduction methods for the SITe 2K and FLI IMG4301E CCD cameras are the same as those described by 
Lee et al. (2007, 2011). The observations of both 2011 and 2012 seasons were obtained with a PIXIS: 2048B CCD camera that offers 
cooling down to $-70\rm ^o$ C. The e2v CCD42-40 chip has 2048$\times$2048 pixels and a pixel size of 13.5 $\mu$m. 
With the conventional IRAF package, we processed the CCD frames to correct for bias level and pixel-to-pixel inhomogeneities of 
quantum efficiency (flat field correction) and applied simple aperture photometry to obtain instrumental magnitudes.

Following the procedure described by Lee et al. (2010), TYC 2827-103-1 ($V\rm_T$=10.99, ($B$--$V$)$\rm_T$=+0.64) and TYC 2827-72-1 
($V\rm_T$=10.98, ($B$--$V$)$\rm_T$=+1.16), imaged on the chip at the same time as the variable (V), were chosen as comparison (C) 
and check (K) stars, respectively. From the 2007 and 2008 seasons, a total of 3,027 individual observations were obtained in 
the three bandpasses (1007 in $B$, 1012 in $V$, and 1008 in $R$) and a sample of them is listed in Table 2. The light curves are 
plotted in the top panel of Figure 1 as the (V$-$C) magnitude differences {\it versus} orbital phase, which was computed according 
to the quadratic ephemeris for our binary model (Model 2) determined later in this article with the Wilson-Devinney synthesis code 
(Wilson \& Devinney 1971, hereafter W-D). The differences ('07-'08) between the two seasons are shown in the middle panel and 
the magnitude differences between the check and comparison stars appear in the bottom panel. The 1$\sigma$-values of the dispersion 
of the (K$-$C) differences are $\pm$0.016 mag, $\pm$0.012 mag, and $\pm$0.013 mag from $B$ to $R$ bandpasses, respectively, 
for the 2007 season and $\pm$0.009 mag, $\pm$0.005 mag, and $\pm$0.005 mag for the 2008 season. Whereas the reference stars were 
evidently constant, the light curves display season-to-season light variability, with evident changes in the phase interval 
encompassing primary eclipse.

\section{LIGHT-CURVE SYNTHESIS AND ABSOLUTE DIMENSIONS}

As shown in Figure 1, our observations display a typical light curve of an overcontact system and a flat bottom at primary minimum, 
indicating that the smaller primary star is totally occulted by the secondary. This would mean that EP And belongs to the W-type 
of W UMa stars. On the contrary, Pribulla et al. (2001) and Manzoori (2012) analyzed their $BV$ light curves and both the AAVSO 
and WASP datasets, respectively, and classified the binary system as a member of the A-type category with the mass ratios of 0.34 
and 0.395. The light curves of 2008 present equal light levels at the quadratures (Max I and Max II) within about 0.002 mag, 
while those of 2007 show the O'Connell effect with Max I brighter than Max II by about 0.014, 0.011, and 0.009 mag for 
the $B$, $V$, and $R$ bandpasses, respectively. The effect is usually interpreted as spot activity on the component stars and 
the seasonal light variations most likely arise from the variability of the spots with time presumably produced by a magnetic dynamo. 

In order to obtain a consistent solution of EP And, we simultaneously solved all avaliable light curves using contact mode 3 of 
the W-D synthesis code and with a weighting scheme identical to that for the eclipsing binary GW Gem (Lee et al. 2009b). Table 3 
lists the light-curve sets for EP And analyzed in this paper and the standard deviations ($\sigma$) of a single observation.
Although the binary parameters have been reported by Pribulla et al. (2001) and Manzoori (2012), their solutions with $q<$1.0 
do not correspond to our high-precision observations of 2008, showing a total eclipse at primary minimum and hence implying $q>$1.0. 
To resolve this confusion, we analyzed the light curves of EP And in a manner similar to that for the overcontact systems AR Boo 
(Lee et al. 2009c) and GW Cep (Lee et al. 2010) using the so-called $q$-search procedure. In the computation of our solutions, 
the surface temperature of the larger, and presumably more massive, star was held fixed at 6,360 K from Flower's (1996) table, 
according to ($B-V$)=$+$0.57$\pm$0.02 given by Terrell et al. (2012) and $E$($B-V$)=$+$0.09 calculated following 
Schlegel et al. (1998). The logarithmic bolometric ($X$, $Y$) and monochromatic ($x$, $y$) limb-darkening coefficients were 
interpolated from the values of van Hamme (1993) in concert with the model atmosphere option. Before the light curves are analyzed, 
the AAVSO times were transformed from JD into HJD. Further, two LTT effects proposed in the following section were applied to 
the observed times of all individual points: HJD$_{\rm new}$=HJD$_{\rm obs}$--($\tau_{3}$+$\tau_{4}$). The quantities shifted by 
the two LTTs are 0.00722$\sim$0.00871 d for 2001, 0.00733$\sim$0.00831 d for 2007, and 0.00026$\sim$0.00052 d for 2008. 
In this paper, we refer to the primary and secondary stars as those being eclipsed at Min I and Min II, respectively. 

Terrell \& Wilson (2005) showed that the mass ratio for a totally-eclipsing overcontact system can be accurately determined from 
a light-curve analysis. Thus, we conducted the detailed $q$-search procedures for both all datasets and only the 2008 light curves, 
permitting no perturbations such as a third light ($l_3$) or a spot. As displayed in Figure 2, the $q$-search results indicate 
a minimum value of the weighted sum of the squared residuals ($\Sigma$) around $q$=2.75. This value corresponds to an occultation 
at primary minimum and indicates that EP And is a W-type overcontact binary. To obtain an unperturbed solution (Model 1), 
we analyzed all light curves by treating the initial value of $q$ as a free parameter. The results are listed in columns (2)--(3)
of Table 4. Then, we reanalyzed the EP And curves by considering a third light ($l_3$) as an additional adjustable parameter 
because it had been suggested by our period study later. New results with the third light source are given as Model 2 in Table 4 
and the light residuals from this binary model are plotted in Figure 3. There is a statistically significant difference in 
the mass ratios between the two models due to the third light effect. In a formal sense, as shown by the entries on the last line 
of the table, the third-light model gives a smaller value of $\Sigma W(O-C)^2$.

As indicated by Figure 3, the model light curves describe the observations of the 2001 and 2008 seasons quite well, but not those 
of the 2007 season. The non-modelled light could be explained by a magnetic cool spot on either of the component stars. Model spots 
were added to fit the small light variations of 2007 by adjusting only the spot and luminosity parameters among the Model 2 parameters. 
Final results are given in Table 5, and the residuals from the cool-spot model on the secondary star are plotted as the plus symbols 
in the middle panels of Figure 3. From these displays, we can see that the spot model does fit the asymmetries in the light maxima 
acceptably but it is difficult to distinguish between the cool-spot models because there is no $\Sigma$ differences among them.
Our light-curve solutions indicate that EP And is a totally-eclipsing W-type overcontact binary with a fill-out factor of about 28 \% 
and with a small temperature difference of 27 K between the components and that $l_3$ contributes 2--3 \% light in all bandpasses.

Absolute dimensions for EP And can be estimated from the photometric solutions (Model 2) in Table 4 and from Harmanec's (1988) relation 
between temperature (spectral type) and stellar mass. We assumed the more massive secondary star to be a normal main-sequence one 
with a spectral type of about F6 and computed the physical properties for the system listed in Table 6. The luminosity ($L$) and 
bolometric magnitudes ($M_{\rm bol}$) were computed by adopting $T_{\rm eff}$$_\odot$=5,780 K and $M_{\rm bol}$$_\odot$=+4.73 
for solar values. For the absolute visual magnitudes ($M_{\rm V}$), we used the bolometric corrections (BCs) appropriate for the temperature 
of each component from the expression between $\log T_{\rm eff}$ and BC given by Torres (2010). Using an apparent visual magnitude of 
$V$=+11.90 (Terrell et al. 2012) at maximum light, the computed light ratio at phase 0.25, and the interstellar absorption of $A_{\rm V}$=0.28, 
we calculated an approximate distance to the system of about 470 pc. In the mass-radius, mass-luminosity, and Hertzsprung-Russell diagrams 
from Hilditch et al. (1988), the locations of both components of EP And conform to the general pattern of overcontact binaries.

\section{ORBITAL PERIOD STUDY}

From the SOAO observations, 29 new times of minimum light and their errors were determined with the weighted means for the timings 
in each filter by using the method of Kwee \& van Woerden (1956). In addition, 61 eclipses were newly derived by us from the WASP data 
and five timings from the AAVSO data. The AAVSO times from Manzoori (2012) may be not HJD but JD. For a period study of EP And, 
we have collected a total of 414 timings (58 photographic plate, 216 visual, 140 photoelectric and CCD) including our measurements. 
All photoelectric and CCD timings are listed in Table 7, wherein the second column gives the HJED (Heliocentric Julian Ephemeris Date) 
timings transformed to the terrestrial time scale (Bastian 2000). Because most earlier timings from Kreiner et al. (2001) were published 
without error information, we calculated the standard deviations of the scatter bands of the timing residuals to provide mean errors 
for the observational methods, as follows: $\pm$0.0101 d for photographic plate, $\pm$0.0076 d for visual, and $\pm$0.0012 d for PE 
and CCD minima. Relative weights were then scaled from the inverse squares of these values consistent with the errors and weights 
for the PE and CCD timings.

The orbital period of EP And was studied for the first time by Qian \& Yuan (2001). From a quadratic least-squares fit, they reported 
a period increase with a rate of $+$1.16$\times$10$^{-7}$ d yr$^{-1}$. Recently, Manzoori (2012) claimed that two periodicities of 
41.2 yr and 11.7 yr, superimposed on the upward parabolic variation, exist in the timing residuals. As the first step for 
ephemeris computations, we applied a periodogram analysis to the complete dataset using the \texttt{PERIOD04} program (Lenz \& Breger 2005). 
As can be seen from Figure 4, two frequencies of $f_1$=0.0000424 cycle d$^{-1}$ and $f_2$=0.00149 cycle d$^{-1}$ were detected 
corresponding to 23,585 d (64.6 yr) and 671 d (1.8 yr), respectively. Thus, the two periods were used to provide an initial guess 
for the Levenberg-Marquart (LM) fitting procedure (Press et al. 1992). The oscillations were assumed to be due to a combination of 
two LTT effects caused by the third and fourth bodies in the system and all times of minimum light were fitted to the following 
two-LTT ephemeris: 
\begin{eqnarray}
C_1 = T_0 + PE + \tau_3 + \tau_4.
\end{eqnarray}
Here, $\tau_{3}$ and $\tau_{4}$ are the LTT due to two additional companions orbiting the eclipsing pair (Irwin 1952, 1959) and each 
includes five parameters ($a_{12}\sin i$, $e$, $\omega$, $n$, $T$). The LM technique was applied to solve for the twelve parameters 
of the ephemeris and the results are listed in columns (2)--(3) of Table 8, together with related quantities. Our absolute dimensions 
presented in Table 6 have been used for these and subsequent calculations.  

The $O$--$C_1$ diagram constructed with the linear terms of the two-LTT ephemeris is plotted in the top panel of Figure 5, where 
the solid and dashed curves represent the full contribution and the $\tau_{3}$ orbit, respectively. The middle panel displays 
the PE and CCD residuals from the complete ephemeris and the bottom panel represents the $\tau_{4}$ orbit. As displayed in the figure, 
all times of minimum light currently agree with the two-LTT ephemeris satisfactorily. The successful fit to the times of minimum light 
with the two LTT orbits tempted us to try to discover a secular term which might be hidden in the two periodic variations. 
Because EP And is in an overcontact configuration with common convective envelope, a parabolic variation should be produced by 
mass transfer between both components and/or by AML due to magnetic stellar wind. Therefore, a more general fit to the times of 
minimum light was made by adding a quadratic term to the two-LTT ephemeris:
\begin{eqnarray}
C_2 = T_0 + PE + A E^2 + \tau_3 + \tau_4.
\end{eqnarray}
The calculations using the LM method converged quickly to yield the results given in columns (4)--(5) of Table 8. 
The $O$--$C_2$ diagram constructed with the linear light elements are drawn at the top of Figure 6 with the solid curve due to 
the sum of the non-linear terms and the dashed parabola due to the quadratic term of equation (2). The second to bottom panels
are plotted in the same sense as in Figure 5. The timing residuals from the full ephemeris appear as $O$--$C_{\rm 2,full}$ 
in the fifth column of Table 7. Figure 7 shows the PE and CCD residuals phased with the $\tau_{4}$ cycle (1.8340 yr) listed 
in Table 8. This ephemeris resulted in a smaller $\chi^2_{\rm red}$=0.981 than the two-LTT ephemeris ($\chi^2_{\rm red}$=1.102). 
Its long-term period ($\tau_4$) is short compared to that of the two-LTT ephemeris by about 21 yr, while the short-term periods 
($\tau_3$) for the two ephemerides are in excellent agreement with each other. 

If it is assumed that the orbits of the two circumbinary objects are coplanar with that of the eclipsing pair of EP And 
($i_{3,4}$=83\fdg3), the masses of the third and fourth bodies are $M_3$=0.25 M$_\odot$ and $M_4$=0.90 M$_\odot$, respectively. 
If they are main-sequence stars, the radii and temperatures are calculated to be $R_3$=0.26 R$_\odot$ and $T_3$=3039 K, and 
$R_4$=0.92 R$_\odot$ and $T_4$=5082 K from the empirical relation of Southworth (2009). The third and fourth bodies would 
contribute about 0.1 \% and 11.6 \%, respectively, to the total bolometric luminosity of the quadruple system. Because 
our light-curve solutions in Tables 4 and 5 detected only $l_3$ of 2--3 \% in all bandpasses, the putative fourth object 
have to be very under-luminous in comparison to the binary components and may be a compact star. Alternatively, it is possible
that the fourth body might be a binary itself. This could reduce the luminosity for the given total mass of 0.90 M$_\odot$
and implies that EP And should be a quintuple system. The semi-amplitude of the expected systemic radial velocity changes of 
the eclipsing pair due to the third and fourth components would be about 1.5 km s$^{-1}$ and 11.8 km s$^{-1}$, respectively. 
Hence, the M-type third companion is very difficult to reveal because of the low systemic velocity change and 
the large orbital period suggested by the LTT model, while the massive circumbinary object might be easily detected with 
high-resolution spectroscopy. On the other hand, the eclipse timing variations might be partly caused by the perturbative effect 
of the fourth component added to the geometrical LTT effect because its LTT period is very short (Borkovits et al. 2003, 2011). 
One such example is IU Aur: \" Ozdemir et al (2003) showed that the third companion with a period of 293.3 d cause the non-negligible, 
dynamical contribution to the $O$--$C$ curve. We computed the semi-amplitude of the fourth-body dynamic perturbation on the motion 
of the overcontact binary to be 0.000013 d and found that its contribution is not significant. 

The positive coefficient of the quadratic term ($A$) listed in Table 8 yields a secular period increase with a rate of 
+5.09$\times$10$^{-8}$ d yr$^{-1}$, corresponding to a fractional period change of +1.39$\times$10$^{-10}$. This value agrees 
well with the value of $+$1.40$\times$10$^{-10}$ calculated with our W-D binary code, independently of the eclipse timings. 
Under the assumption of conservative mass transfer, this gives a continuous mass transfer from the less massive primary to 
the secondary component at a modest rate of 3.41$\times$10$^{-8}$ M$_\odot$ yr$^{-1}$. The observed value is small by a factor of 
about 50\% compared with the predicted rate of 6.77$\times$10$^{-8}$ M$_\odot$ yr$^{-1}$ calculated by assuming that the primary 
transfers its present mass to the secondary on a thermal time scale. Thus, the possible explanation of the parabolic variation 
might be some combination of non-conservative mass transfer and AML due to magnetic braking.

\section{DISCUSSION AND CONCLUSIONS}

In this article, we presented and analyzed new long-term CCD observations of EP And, together with historical data collected 
from the literature. The light curves display a total eclipse at primary minimum and season-to-season light variability. 
The asymmetric light curves in 2007 were modeled by a magnetic cool spot on either of the component stars. Our detailed study 
of the light curves and the orbital period represent EP And to be a quadruple (or a quintuple) system with a W-type overcontact binary. 
Because the period is increasing and the mass is transferring from the primary star to the more massive secondary, the eclipsing pair 
may presently be in an expanding state evolving from a overcontact to a non-contact configuration as it undergoes thermal relaxation 
oscillations (Lucy 1976; Lucy \& Wilson 1979).
 
In principle, the periodic variations in the eclipse timing residuals can be expected because of stellar activity variations 
of a magnetically active star, as was initially proposed by Applegate (1992) and later modified by Lanza et al. (1998). With 
the modulation periods ($P_{3,4}$) and amplitudes ($K_{3,4}$) listed in columns (4)--(5) of Table 8, the model parameters 
for each cycle were calculated from the Applegate formulae and are listed in Table 9, where the rms luminosity changes 
($\Delta m_{\rm rms}$) converted to magnitude scale were obtained with equation (4) in the paper of Kim et al. (1997). 
In the table, the primary component with 1.2 L$_\odot$ and the secondary with 2.7 L$_\odot$ exhibit 
the predicted luminosity variations of 55.4 L$_\odot$ and 34.8 L$_\odot$, respectively, for the short-term cycle. The variations 
of the gravitational quadrupole moment ($\Delta Q$) for the long-term cycle are two orders of magnitude smaller than typical values 
of $10^{51}-10^{52}$ for close binaries (Lanza \& Rodono 1999). Moreover, it is difficult for the model to produce perfectly 
smooth and tilted periodic components in the eclipse timing variation. These suggest that Applegate mechanism cannot explain 
the observed period modulations of EP And. On the other hand, a single periodic variation could be attributed to the rotation 
of the apsidal line of the binary orbit due to tidal forces between the two binary components. However, our light-curve analysis 
suggest that the binary orbit is circular which in turn excludes timing variations from apsidal precession.

As can be seen in Figures 5--7, all times of minimum light agree quite well by interpreting the observed LTT signal as caused 
by a third and fourth body surrounding the eclipsing pair. The possible existence of the circumbinary objects is consistent 
with the suggestion of Pribulla \& Rucinski (2006) that most W UMa systems exist in multiple systems. 
We have carried out a stability study for the orbit parameters in Table 8, under the assumption that the eclipsing pair 
can be replaced by a single massive object with mass equal to the combined mass of the two binary components. In general, 
our methodology on assessing the orbital stability is similar to the work presented in Hinse et al. (2012) who considers 
the stability of a similar system (SZ Herculis) of two circumbinary M-type companions (Lee et al. 2012). 
The dynamical stability test suggests that the two proposed companions are on highly unstable orbits. In order to reconcile 
the apparent contradiction we are therefore left with two options. Either i) we discard our two-companion interpretation 
or ii) our LTT model is in lack of important physics that is not included in the present analysis/model. In light of 
the above given arguments we do not favour to discard the two companion interpretation as presented in this work. 
Of the possible causes of the periodic variations, the Applegate effect and apsidal motion can be ruled out. 
The most reasonable explanation of both cycles is a pair of the LTT effects driven by the presence of circumbinary companions. 
In future work we plan a re-analysis of the data with an improved model that formulates the LTT effect in Jacobi coordinates 
and includes mutual gravitational interactions. Mutual interactions between the companions is a highly non-linear process and 
could result in a significant different orbital architecture consistent with the observed timing data. The two-Kepler assumption 
might be inadequate when larger companion masses are involved. This system is an obvious candidate for 
future photometric follow-up programs for further characterisation (Pribulla et al. 2012). 

Assuming our LTT interpretation is correct, then the outer components may have played an important role in the formation and 
evolution of the inner eclipsing pair, which would cause it to evolve into an overcontact configuration by AML via magnetic braking 
and ultimately to coalesce into a single rapid-rotating star. High-resolution photometry and spectroscopy will help to identify and 
understand the orbital period variation of the binary system and to determine the absolute parameters and evolutionary status of the 
multiple system better than is possible with photometry alone.

\acknowledgments{ }

We would like to thank the staff of the Sobaeksan Optical Astronomy Observatory for assistance during our observations. 
We appreciate the careful reading and valuable comments of the reviewer Dirk Terrell. This research has made use of 
the Simbad database maintained at CDS, Strasbourg, France. We have used data from the AAVSO International Database and 
the WASP public archive in this research. The WASP consortium comprises of the University of Cambridge, Keele University, 
University of Leicester, The Open University, The Queen's University Belfast, St. Andrews University and the Isaac Newton Group. 
Funding for WASP comes from the consortium universities and from the UK's Science and Technology Facilities Council. This work was 
supported by the KASI (Korea Astronomy and Space Science Institute) grant 2013-9-400-00. T.C.H. acknowledges financial support from 
the Korea Research Council for Fundamental Science and Technology (KRCF) through the Young Research Scientist Fellowship Program.

\clearpage
\begin{figure}
 \includegraphics[]{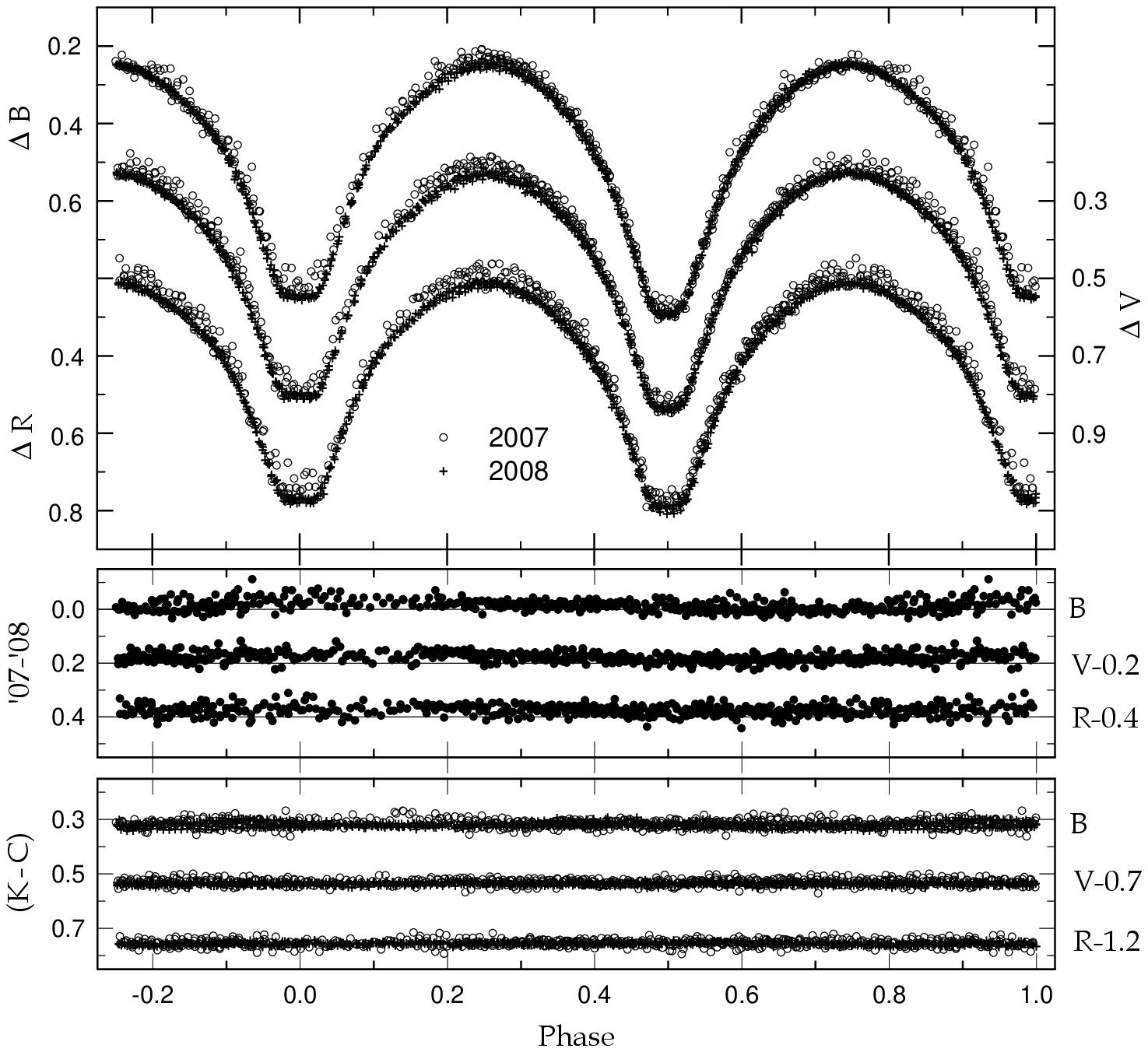}
 \caption{The top panel displays our light curves of EP And in the $B$, $V$, and $R$ bandpasses. The differences 
 between the two seasons are shown in the middle panel and the magnitude difference between the check and comparison stars 
 in the lower panel. The open circles and plus symbols in the top and bottom panels are the individual measures of 
 the 2007 and 2008 seasons, respectively. The solid lines in the middle panel refer to values of 0.0 mag}
 \label{Fig1}
\end{figure}

\begin{figure}
 \includegraphics[]{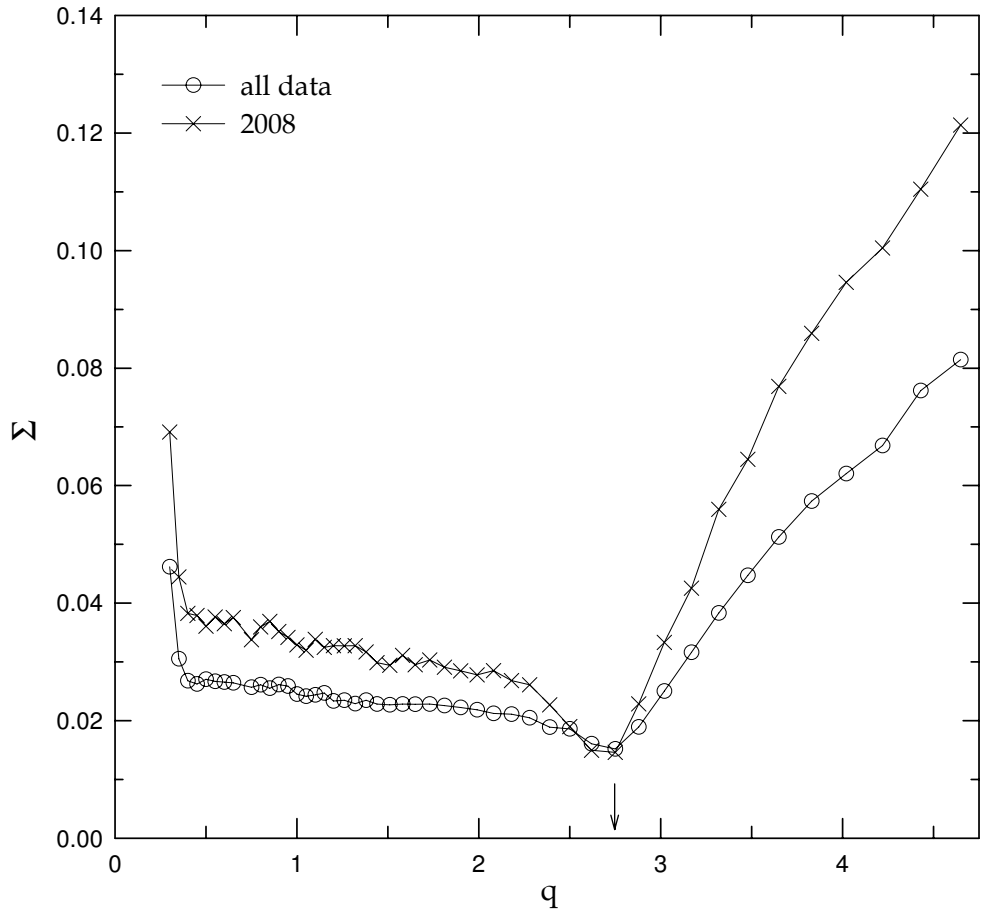}
 \caption{Behavior of $\Sigma$ as a function of mass ratio $q$ for all datasets and the 2008 light curves of EP And. Both indicate 
 a minimum value near $q$=2.75. }
\label{Fig2}
\end{figure}

\begin{figure}
 \includegraphics[]{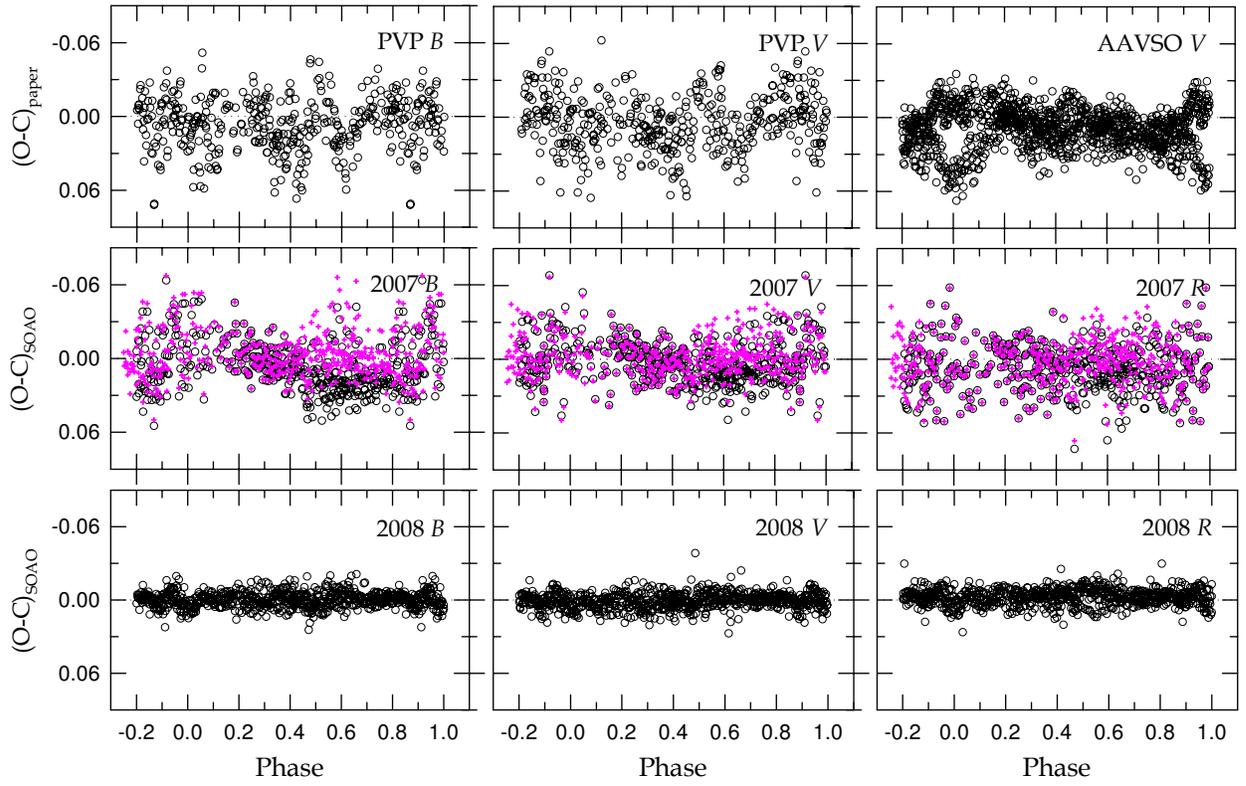}
 \caption{The magnitude residuals from the binary model with a third light in columns (4)--(5) of Table 4. The plus symbols in middle panels
 represent the residuals from the cool-spot model on the more massive secondary star for the 2007 light curves.
 } 
\label{Fig3}
\end{figure}

\begin{figure}
 \includegraphics[]{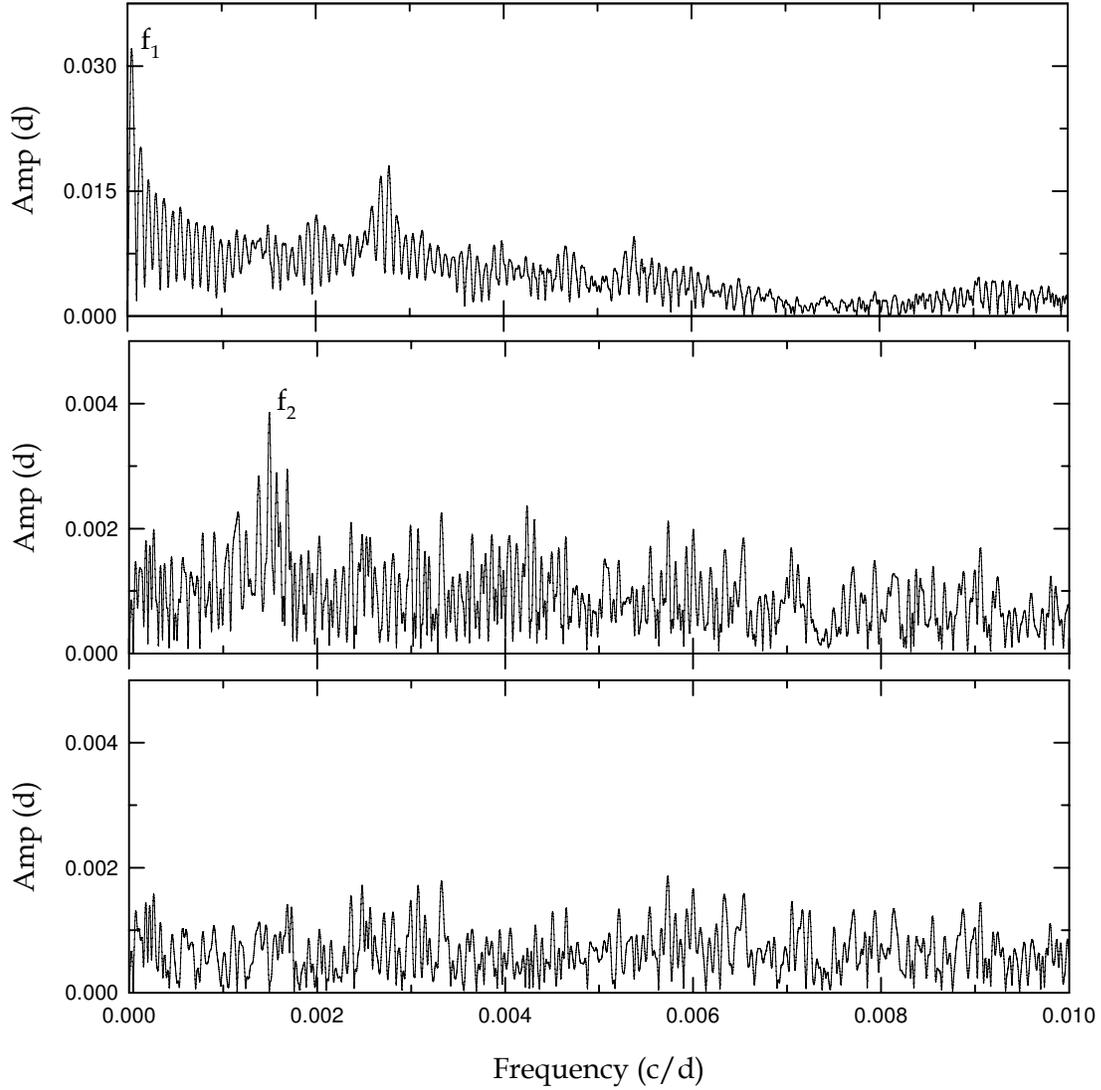}
 \caption{Periodogram from the \texttt{PERIOD04} formalism for all observed timings. As a result of the successive pre-whitening procedures, 
 two frequencies of $f_1$=0.0000424 cycle d$^{-1}$ and $f_2$=0.00149 cycle d$^{-1}$ are detected and these become periods of 23,585 d and 671 d.}
 \label{Fig4}
\end{figure}

\begin{figure}
 \includegraphics[]{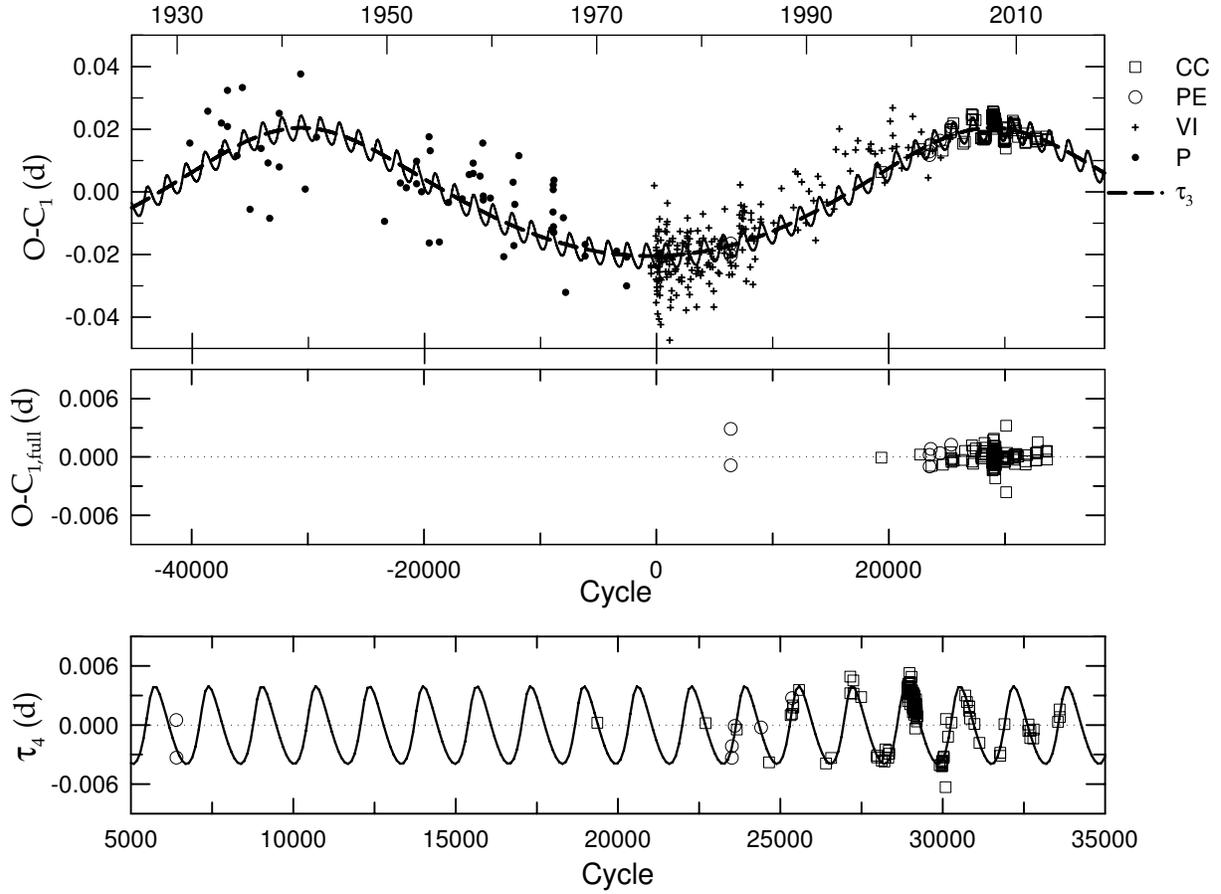}
 \caption{$O$--$C$ diagram of EP And with respect to the linear terms of equation (1). In the top panel, the two-LTT ephemeris is 
 drawn as the solid curve and the dashed curve represents the $\tau_3$ orbit. The middle panel displays the PE and CCD residuals 
 from the complete ephemeris and the bottom panel represents the $\tau_4$ orbit. CC, PE, VI, and P denote CCD, photoelectric, visual, 
 and photographic plate minima, respectively. }
\label{Fig5}
\end{figure}

\begin{figure}
 \includegraphics[]{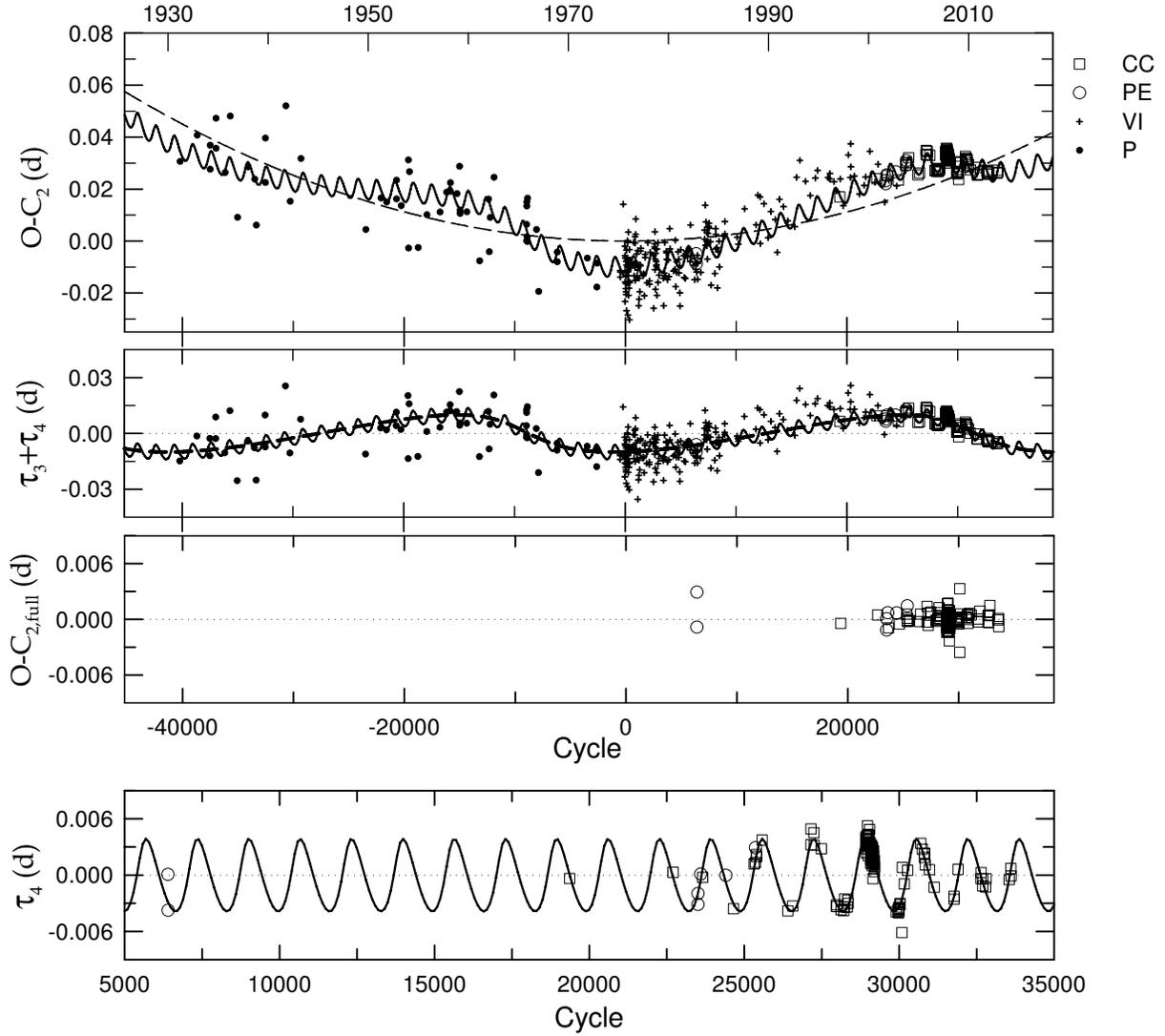}
 \caption{In the top panel the $O$--$C$ diagram of EP And constructed with the linear terms of the quadratic {\it plus} two-LTT ephemeris. 
 The full ephemeris is drawn as the solid curve and the dashed parabola is due to only the quadratic term of equation (2). The second and 
 third panels display the residuals ($\tau_3$+$\tau_4$) from the quadratic term and the PE and CCD residuals from the complete ephemeris, 
 respectively. The bottom panel represents the residuals $\tau_4$ from the quadratic term {\it plus} $\tau_3$, respectively. }
\label{Fig6}
\end{figure}

\begin{figure}
 \includegraphics[]{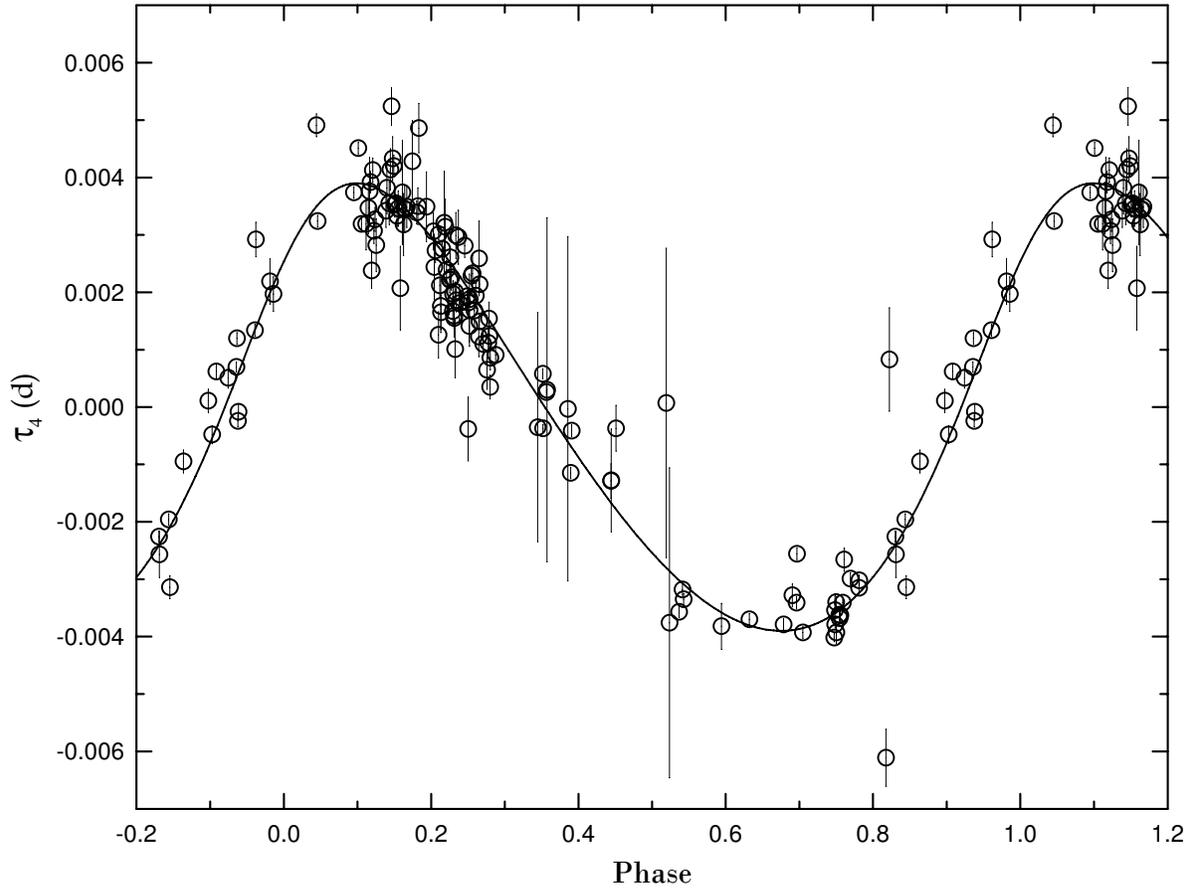}
 \caption{Photoelectric and CCD residuals from the bottom panel of Figure 6 phased with the $\tau_4$ cycle (1.8340 yr) of 
 the quadratic {\it plus} two-LTT ephemeris listed in Table 8. }
\label{Fig7}
\end{figure}

\clearpage
\begin{deluxetable}{llccccc}
\tabletypesize{\small}
\tablewidth{0pt} 
\tablecaption{Summary of the SOAO observations of EP And.}
\tablehead{
\colhead{Season}   &  \colhead{Observing Interval}   &  \colhead{$N_{\rm night}$}   &  \colhead{CCD Type}  &  \colhead{FOV (arcmin$^2$)}  &  \colhead{Note}   &  \colhead{Ref.\rm $^a$}
}
\startdata
2007               &  Oct. 26 $\sim$ Nov. 24         &  8                           &  SITe 2K             &  20.5$\times$20.5            &  light curve      &  (1)          \\  
2008               &  Sep. 30 $\sim$ Nov. 04         &  10                          &  SITe 2K             &  20.5$\times$20.5            &  light curve      &  (1)          \\  
2009               &  Feb. 08                        &  1                           &  SITe 2K             &  20.5$\times$20.5            &  eclipse timing   &  (1)          \\  
2009               &  Sep. 17 $\sim$ Nov. 21         &  3                           &  FLI IMG4301E 2K     &  20.9$\times$20.9            &  eclipse timing   &  (2)          \\  
2010               &  Nov. 29                        &  1                           &  FLI IMG4301E 2K     &  20.9$\times$20.9            &  eclipse timing   &  (2)          \\  
2011               &  Sep. 22 $\sim$ Oct. 18         &  4                           &  PIXIS: 2048B        &  17.6$\times$17.6            &  eclipse timing   &               \\  
2012               &  Sep. 25 $\sim$ Oct. 19         &  3                           &  PIXIS: 2048B        &  17.6$\times$17.6            &  eclipse timing   &               \\  
\enddata
\tablenotetext{a}{(1) Lee et al. (2007); (2) Lee et al. (2011).}
\end{deluxetable}

\begin{deluxetable}{crcrcr}
\tablewidth{0pt} 
\tablecaption{CCD photometric data of EP And in 2007 and 2008.}
\tablehead{
\colhead{HJD} & \colhead{$\Delta B$} & \colhead{HJD} & \colhead{$\Delta V$} & \colhead{HJD} & \colhead{$\Delta R$} 
}
\startdata
2,454,400.20520  &  0.337   &   2,454,400.20669  &  0.283   &   2,454,400.20786  &  0.249     \\
2,454,400.20931  &  0.325   &   2,454,400.21080  &  0.271   &   2,454,400.21197  &  0.244     \\
2,454,400.21341  &  0.303   &   2,454,400.21491  &  0.256   &   2,454,400.21608  &  0.229     \\
2,454,400.21752  &  0.296   &   2,454,400.21902  &  0.235   &   2,454,400.22019  &  0.209     \\
2,454,400.22163  &  0.282   &   2,454,400.22313  &  0.215   &   2,454,400.22430  &  0.203     \\
2,454,400.22574  &  0.256   &   2,454,400.22723  &  0.220   &   2,454,400.22841  &  0.189     \\
2,454,400.22985  &  0.245   &   2,454,400.23134  &  0.208   &   2,454,400.23252  &  0.190     \\
2,454,400.23396  &  0.251   &   2,454,400.23545  &  0.212   &   2,454,400.23663  &  0.184     \\
2,454,400.23807  &  0.246   &   2,454,400.23956  &  0.197   &   2,454,400.24074  &  0.169     \\
2,454,400.24218  &  0.222   &   2,454,400.24367  &  0.196   &   2,454,400.24484  &  0.181     \\
\enddata
\tablecomments{This table is available in its entirety in machine-readable and Virtual Observatory (VO) forms 
in the online journal. A portion is shown here for guidance regarding its form and content.}
\end{deluxetable}

\begin{deluxetable}{lccc}
\tablewidth{0pt}
\tablecaption{Light-curve sets for EP And.}
\tablehead{
\colhead{Reference}   & \colhead{Season} & \colhead{Filter}    & \colhead{$\sigma$$\rm ^a$} }
\startdata
PVP$\rm ^b$           & 2001             & $B$                 & 0.0162            \\
                      &                  & $V$                 & 0.0169            \\
AAVSO$\rm ^c$         & 2007             & $V$                 & 0.0122            \\					  
SOAO                  & 2007             & $B$                 & 0.0136            \\
                      &                  & $V$                 & 0.0122            \\
                      &                  & $R$                 & 0.0144            \\
                      & 2008             & $B$                 & 0.0048            \\
                      &                  & $V$                 & 0.0051            \\
                      &                  & $R$                 & 0.0051            \\
\enddata
\tablenotetext{a}{In units of total light at phase 0.25.}
\tablenotetext{b}{Pribulla et al. (2001).}
\tablenotetext{c}{Manzoori (2012).}
\end{deluxetable}

\begin{deluxetable}{lccccc}
\tabletypesize{\scriptsize} %{\small}
\tablewidth{0pt} 
\tablecaption{EP And parameters determined by analyzing simultaneously all curves$\rm ^a$.}
\tablehead{
\colhead{Parameter}                      & \multicolumn{2}{c}{Model 1}                      && \multicolumn{2}{c}{Model 2}                      \\ [1.0mm] \cline{2-3} \cline{5-6} \\[-2.0ex]
                                         & \colhead{Primary} & \colhead{Secondary}          && \colhead{Primary} & \colhead{Secondary}              
}
\startdata                                                                                                                                          
$T_0$ (HJD)                              & \multicolumn{2}{c}{2,442,638.5230(23)}           && \multicolumn{2}{c}{2,442,638.5230(22)}           \\ 
$P$ (d)                                  & \multicolumn{2}{c}{0.40410886$\pm$(17)}          && \multicolumn{2}{c}{0.40410886$\pm$(16)}          \\ 
d$P$/d$t$                                & \multicolumn{2}{c}{1.40(15)$\times$10$^{-10}$}   && \multicolumn{2}{c}{1.40(15)$\times$10$^{-10}$}   \\
$q$                                      & \multicolumn{2}{c}{2.7198(21)}                   && \multicolumn{2}{c}{2.5784(9)}                    \\
$i$ (deg)                                & \multicolumn{2}{c}{83.01(7)}                     && \multicolumn{2}{c}{83.31(10)}                    \\
$T$ (K)                                  & 6,387(2)          & 6,360                        && 6,387(2)          & 6,360                        \\
$\Omega$                                 & 6.0759(38)        & 6.0759                       && 5.8809(25)        & 5.8809                       \\
$\Omega_{\rm in}$                        & \multicolumn{2}{c}{6.2426}                       && \multicolumn{2}{c}{6.0518}                       \\
$A$                                      & \multicolumn{2}{c}{0.5}                          && \multicolumn{2}{c}{0.5}                          \\
$g$                                      & \multicolumn{2}{c}{0.32}                         && \multicolumn{2}{c}{0.32}                         \\
$X$, $Y$                                 & \multicolumn{2}{c}{0.641, 0.235}                 && \multicolumn{2}{c}{0.641, 0.235}                 \\
$x_{B}$, $y_{B}$                         & \multicolumn{2}{c}{0.811, 0.223}                 && \multicolumn{2}{c}{0.811, 0.223}                 \\
$x_{V}$, $y_{V}$                         & \multicolumn{2}{c}{0.719, 0.272}                 && \multicolumn{2}{c}{0.719, 0.272}                 \\
$x_{R}$, $y_{R}$                         & \multicolumn{2}{c}{0.646, 0.282}                 && \multicolumn{2}{c}{0.646, 0.282}                 \\
$l$/($l_{1}$+$l_{2}$+$l_{3}$){$_{B}$}    & 0.2972(5)         & 0.7028                       && 0.2992(17)        & 0.6739                       \\
$l$/($l_{1}$+$l_{2}$+$l_{3}$){$_{V}$}    & 0.2957(4)         & 0.7043                       && 0.2962(18)        & 0.6718                       \\
$l$/($l_{1}$+$l_{2}$+$l_{3}$){$_{V}$}    & 0.2957(3)         & 0.7043                       && 0.2983(8)         & 0.6767                       \\
$l$/($l_{1}$+$l_{2}$+$l_{3}$){$_{B}$}    & 0.2972(5)         & 0.7028                       && 0.2990(14)        & 0.6734                       \\
$l$/($l_{1}$+$l_{2}$+$l_{3}$){$_{V}$}    & 0.2957(4)         & 0.7043                       && 0.3015(13)        & 0.6839                       \\
$l$/($l_{1}$+$l_{2}$+$l_{3}$){$_{R}$}    & 0.2950(4)         & 0.7050                       && 0.2978(14)        & 0.6780                       \\
$l$/($l_{1}$+$l_{2}$+$l_{3}$){$_{B}$}    & 0.2972(3)         & 0.7028                       && 0.3020(5)         & 0.6801                       \\
$l$/($l_{1}$+$l_{2}$+$l_{3}$){$_{V}$}    & 0.2957(3)         & 0.7043                       && 0.2998(5)         & 0.6800                       \\
$l$/($l_{1}$+$l_{2}$+$l_{3}$){$_{R}$}    & 0.2950(3)         & 0.7050                       && 0.2977(5)         & 0.6776                       \\
{\it $l_{3B}$$\rm ^b$}                   &  \multicolumn{2}{c}{0.0}                         && \multicolumn{2}{c}{0.0269(43)}                   \\
{\it $l_{3V}$$\rm ^b$}                   &  \multicolumn{2}{c}{0.0}                         && \multicolumn{2}{c}{0.0320(47)}                   \\
{\it $l_{3V}$$\rm ^b$}                   &  \multicolumn{2}{c}{0.0}                         && \multicolumn{2}{c}{0.0250(22)}                   \\
{\it $l_{3B}$$\rm ^b$}                   &  \multicolumn{2}{c}{0.0}                         && \multicolumn{2}{c}{0.0276(34)}                   \\
{\it $l_{3V}$$\rm ^b$}                   &  \multicolumn{2}{c}{0.0}                         && \multicolumn{2}{c}{0.0146(33)}                   \\
{\it $l_{3R}$$\rm ^b$}                   &  \multicolumn{2}{c}{0.0}                         && \multicolumn{2}{c}{0.0242(39)}                   \\
{\it $l_{3B}$$\rm ^b$}                   &  \multicolumn{2}{c}{0.0}                         && \multicolumn{2}{c}{0.0179(14)}                   \\
{\it $l_{3V}$$\rm ^b$}                   &  \multicolumn{2}{c}{0.0}                         && \multicolumn{2}{c}{0.0202(14)}                   \\
{\it $l_{3R}$$\rm ^b$}                   &  \multicolumn{2}{c}{0.0}                         && \multicolumn{2}{c}{0.0247(14)}                   \\
$r$ (pole)                               & 0.2888(3)         & 0.4515(3)                    && 0.2935(2)         & 0.4478(2)                    \\
$r$ (side)                               & 0.3028(4)         & 0.4859(4)                    && 0.3080(3)         & 0.4816(3)                    \\
$r$ (back)                               & 0.3458(8)         & 0.5168(5)                    && 0.3514(5)         & 0.5131(3)                    \\
$r$ (volume)$\rm ^c$                     & 0.3143            & 0.4865                       && 0.3195            & 0.4827                       \\
$\Sigma W(O-C)^2$                        & \multicolumn{2}{c}{0.0128}                       && \multicolumn{2}{c}{0.0125}     
\tablenotetext{a}{Luminosity parameters are listed in the same order as entries in Table 3.}
\tablenotetext{b}{Value at 0.25 phase.}
\tablenotetext{c}{Mean volume radius.}
\enddata
\end{deluxetable}

\begin{deluxetable}{lccc}
\tablewidth{0pt}
\tablecaption{Spot and luminosity parameters for the 2007 light curves.}
\tablehead{
\colhead{Parameter}                      &  \multicolumn{2}{c}{Cool-Spot Model}                            \\ [1.0mm] \cline{2-3} \\[-2.0ex]
                                         &  \colhead{on Primary}          &  \colhead{on Secondary}          
}
\startdata 
Colatitude (deg)                         &  66.1$\pm$2.5                  &  52.6$\pm$2.3                  \\
Longitude (deg)                          &  130.8$\pm$2.3                 &  307.6$\pm$2.8                 \\
Radius (deg)                             &  16.92$\pm$0.47                &  16.06$\pm$0.83                \\
$T$$\rm _{spot}$/$T$$\rm _{local}$       &  0.867$\pm$0.014               &  0.925$\pm$0.008               \\[2.0mm]
$l_1$/($l_{1}$+$l_{2}$+$l_{3}$){$_{B}$}  &  0.3018$\pm$0.0018             &  0.3017$\pm$0.0018             \\
$l_1$/($l_{1}$+$l_{2}$+$l_{3}$){$_{V}$}  &  0.2998$\pm$0.0016             &  0.2998$\pm$0.0016             \\
$l_1$/($l_{1}$+$l_{2}$+$l_{3}$){$_{R}$}  &  0.2976$\pm$0.0018             &  0.2976$\pm$0.0018             \\
{\it $l_{3B}$$\rm ^a$}                   &  0.0184$\pm$0.0040             &  0.0187$\pm$0.0039            \\
{\it $l_{3V}$$\rm ^a$}                   &  0.0201$\pm$0.0036             &  0.0202$\pm$0.0035             \\
{\it $l_{3R}$$\rm ^a$}                   &  0.0248$\pm$0.0040             &  0.0248$\pm$0.0039             \\
$\Sigma W(O-C)^2$                        &  0.0122                        &  0.0122     
\enddata
\tablenotetext{a}{Value at 0.25 phase.}
\end{deluxetable}

\begin{deluxetable}{lcc}
\tablewidth{0pt} 
\tablecaption{Absolute Parameters for EP And.}
\tablehead{
\colhead{Parameter}              & \colhead{Primary} & \colhead{Secondary}
}
\startdata
$M$/$M_\odot$                    &  0.50             &  1.30            \\
$R$/$R_\odot$                    &  0.89             &  1.35            \\
$\log$ $g$ (cgs)                 &  4.24             &  4.29            \\
$\log$ $\rho$/$\rho_\odot$       &  $-$0.15          &  $-$0.28         \\
$T$ (K)                          &  6,387            &  6,360           \\
$L$/$L_\odot$                    &  1.19             &  2.67            \\
$M_{\rm bol}$ (mag)              &  $+$4.54          &  $+$3.66         \\
BC (mag)                         &  0.00             &  0.00            \\
$M_{V}$ (mag)                    &  $+$4.54          &  $+$3.66         \\
Distance (pc)                    &  \multicolumn{2}{c}{470}             \\
\enddata
\end{deluxetable}

\begin{deluxetable}{lllrrcl}
\tablewidth{0pt} 
\tablecaption{Observed PE and CCD times of minimum light for EP And.}
\tablehead{
\colhead{HJD} & \colhead{HJED$\rm ^a$} & Error & \colhead{Epoch} & \colhead{$O$--$C_{\rm 2,full}$} & \colhead{Min} & \colhead{References}  \\
\colhead{(2,400,000+)} & \colhead{(2,400,000+)} & & & & &  }
\startdata
45,221.583    & 45,221.58362  &               &   6392.0  &  $+$0.00292  &  I   &  Hoffmann (1983)               \\
45,224.610    & 45,224.61062  &               &   6399.5  &  $-$0.00085  &  II  &  Hoffmann (1983)               \\
50,463.300    & 50,463.30072  & $\pm$0.002    &  19363.0  &  $-$0.00043  &  I   &  Diethelm (1997)               \\
51,811.616    & 51,811.61674  & $\pm$0.003    &  22699.5  &  $+$0.00048  &  II  &  Paschke (2001)                \\
52,137.5289   & 52,137.52964  & $\pm$0.0001   &  23506.0  &  $+$0.00006  &  I   &  Pribulla et al. (2001)        \\
52,138.5380   & 52,138.53874  & $\pm$0.0002   &  23508.5  &  $-$0.00115  &  II  &  Pribulla et al. (2001)        \\
52,173.4968   & 52,173.49754  & $\pm$0.0002   &  23595.0  &  $+$0.00070  &  I   &  Pribulla et al. (2001)        \\
52,200.3698   & 52,200.37054  & $\pm$0.0001   &  23661.5  &  $-$0.00092  &  II  &  Pribulla et al. (2002)        \\
52,500.422    & 52,500.42274  & $\pm$0.003    &  24404.0  &  $+$0.00069  &  I   &  Locher (2002)                 \\
52,601.6481   & 52,601.64884  & $\pm$0.0001   &  24654.5  &  $-$0.00051  &  II  &  Nelson (2003)                 \\
52,869.3759   & 52,869.37664  & $\pm$0.0001   &  25317.0  &  $+$0.00056  &  I   &  Kotkova \& Wolf (2006)        \\
52,885.7425   & 52,885.74324  & $\pm$0.0001   &  25357.5  &  $-$0.00009  &  II  &  Nelson (2004)                 \\
52,886.5523   & 52,886.55304  & $\pm$0.0003   &  25359.5  &  $+$0.00145  &  II  &  H\"ubscher (2005)             \\
52,899.4831   & 52,899.48384  & $\pm$0.0004   &  25391.5  &  $+$0.00012  &  II  &  Pejcha (2005)                 \\
52,902.5137   & 52,902.51444  & $\pm$0.0003   &  25399.0  &  $-$0.00024  &  I   &  Pejcha (2005)                 \\
52,975.6594   & 52,975.66014  & $\pm$0.00010  &  25580.0  &  $-$0.00012  &  I   &  Samolyk (2011)                \\
53,310.2549   & 53,310.25564  & $\pm$0.0004   &  26408.0  &  $-$0.00023  &  I   &  Kim et al. (2006)             \\
53,374.5089   & 53,374.50964  & $\pm$0.00020  &  26567.0  &  $+$0.00056  &  I   &  Samolyk (2011)                \\
53,611.5274   & 53,611.52814  & $\pm$0.0002   &  27153.5  &  $+$0.00140  &  II  &  Zejda et al. (2006)           \\
53,612.5360   & 53,612.53674  & $\pm$0.0001   &  27156.0  &  $-$0.00030  &  I   &  H\"ubscher et al. (2006)      \\
53,649.5133   & 53,649.51404  & $\pm$0.0001   &  27247.5  &  $+$0.00066  &  II  &  Brat et al. (2007)            \\
53,652.5428   & 53,652.54354  & $\pm$0.0001   &  27255.0  &  $-$0.00066  &  I   &  Biro et al. (2006)            \\
53,746.2958   & 53,746.29655  & $\pm$0.0002   &  27487.0  &  $+$0.00078  &  I   &  H\"ubscher et al. (2006)      \\
53,944.5054   & 53,944.50615  & $\pm$0.0001   &  27977.5  &  $-$0.00007  &  II  &  Parimucha et al. (2007)       \\
53,945.5155   & 53,945.51625  & $\pm$0.0001   &  27980.0  &  $-$0.00022  &  I   &  Parimucha et al. (2007)       \\
54,005.3233   & 54,005.32405  & $\pm$0.0001   &  28128.0  &  $+$0.00010  &  I   &  Parimucha et al. (2007)       \\
54,036.4396   & 54,036.44035  & $\pm$0.0001   &  28205.0  &  $+$0.00007  &  I   &  Csizmadia et al. (2006)       \\
54,048.3612   & 54,048.36195  & $\pm$0.0001   &  28234.5  &  $+$0.00042  &  II  &  Biro et (2007)                \\
54,048.5641   & 54,048.56485  & $\pm$0.0001   &  28235.0  &  $+$0.00126  &  I   &  Biro et (2007)                \\
54,084.3269   & 54,084.32765  & $\pm$0.0001   &  28323.5  &  $+$0.00010  &  II  &  Brat et al. (2007)            \\
54,091.6016   & 54,091.60235  & $\pm$0.0002   &  28341.5  &  $+$0.00073  &  II  &  Nelson (2007)                 \\
54,097.2588   & 54,097.25955  & $\pm$0.0001   &  28355.5  &  $+$0.00031  &  II  &  Brat et al. (2007)            \\
54,326.59669  & 54,326.59744  & $\pm$0.00045  &  28923.0  &  $-$0.00064  &  I   &  This paper (WASP)             \\
54,328.61751  & 54,328.61826  & $\pm$0.00045  &  28928.0  &  $-$0.00035  &  I   &  This paper (WASP)             \\
54,329.62808  & 54,329.62883  & $\pm$0.00059  &  28930.5  &  $-$0.00004  &  II  &  This paper (WASP)             \\
54,330.63851  & 54,330.63926  & $\pm$0.00019  &  28933.0  &  $+$0.00012  &  I   &  This paper (WASP)             \\
54,331.64724  & 54,331.64799  & $\pm$0.00031  &  28935.5  &  $-$0.00141  &  II  &  This paper (WASP)             \\
54,332.65926  & 54,332.66001  & $\pm$0.00021  &  28938.0  &  $+$0.00035  &  I   &  This paper (WASP)             \\
54,333.66847  & 54,333.66922  & $\pm$0.00022  &  28940.5  &  $-$0.00071  &  II  &  This paper (WASP)             \\
54,334.67894  & 54,334.67969  & $\pm$0.00021  &  28943.0  &  $-$0.00050  &  I   &  This paper (WASP)             \\
54,335.68876  & 54,335.68951  & $\pm$0.00046  &  28945.5  &  $-$0.00094  &  II  &  This paper (WASP)             \\
54,344.57975  & 54,344.58050  & $\pm$0.00029  &  28967.5  &  $-$0.00023  &  II  &  This paper (WASP)             \\
54,345.59042  & 54,345.59117  & $\pm$0.00045  &  28970.0  &  $+$0.00018  &  I   &  This paper (WASP)             \\
54,347.61070  & 54,347.61145  & $\pm$0.00037  &  28975.0  &  $-$0.00006  &  I   &  This paper (WASP)             \\
54,348.62156  & 54,348.62231  & $\pm$0.00036  &  28977.5  &  $+$0.00055  &  II  &  This paper (WASP)             \\
54,349.63293  & 54,349.63368  & $\pm$0.00033  &  28980.0  &  $+$0.00166  &  I   &  This paper (WASP)             \\
54,350.64229  & 54,350.64304  & $\pm$0.00038  &  28982.5  &  $+$0.00077  &  II  &  This paper (WASP)             \\
54,351.65243  & 54,351.65318  & $\pm$0.00019  &  28985.0  &  $+$0.00065  &  I   &  This paper (WASP)             \\
54,352.66203  & 54,352.66278  & $\pm$0.00022  &  28987.5  &  $-$0.00000  &  II  &  This paper (WASP)             \\
54,353.67232  & 54,353.67307  & $\pm$0.00008  &  28990.0  &  $+$0.00003  &  I   &  This paper (WASP)             \\
54,354.68238  & 54,354.68313  & $\pm$0.00018  &  28992.5  &  $-$0.00016  &  II  &  This paper (WASP)             \\
54,355.69278  & 54,355.69353  & $\pm$0.00031  &  28995.0  &  $-$0.00002  &  I   &  This paper (WASP)             \\
54,357.71193  & 54,357.71268  & $\pm$0.00073  &  29000.0  &  $-$0.00137  &  I   &  This paper (WASP)             \\
54,359.53209  & 54,359.53284  & $\pm$0.00091  &  29004.5  &  $+$0.00033  &  II  &  This paper (WASP)             \\
54,360.54180  & 54,360.54255  & $\pm$0.00054  &  29007.0  &  $-$0.00021  &  I   &  This paper (WASP)             \\
54,361.55234  & 54,361.55309  & $\pm$0.00041  &  29009.5  &  $+$0.00008  &  II  &  This paper (WASP)             \\
54,362.56260  & 54,362.56335  & $\pm$0.00011  &  29012.0  &  $+$0.00008  &  I   &  This paper (WASP)             \\
54,363.57292  & 54,363.57367  & $\pm$0.00036  &  29014.5  &  $+$0.00015  &  II  &  This paper (WASP)             \\
54,368.62507  & 54,368.62582  & $\pm$0.00071  &  29027.0  &  $+$0.00105  &  I   &  This paper (WASP)             \\
54,373.67565  & 54,373.67640  & $\pm$0.00032  &  29039.5  &  $+$0.00038  &  II  &  This paper (WASP)             \\
54,374.48522  & 54,374.48597  & $\pm$0.00043  &  29041.5  &  $+$0.00175  &  II  &  This paper (WASP)             \\
54,381.55575  & 54,381.55650  & $\pm$0.00061  &  29059.0  &  $+$0.00054  &  I   &  This paper (WASP)             \\
54,387.61695  & 54,387.61770  & $\pm$0.00032  &  29074.0  &  $+$0.00026  &  I   &  This paper (WASP)             \\
54,388.62660  & 54,388.62735  & $\pm$0.00057  &  29076.5  &  $-$0.00033  &  II  &  This paper (WASP)             \\
54,389.63716  & 54,389.63791  & $\pm$0.00021  &  29079.0  &  $-$0.00002  &  I   &  This paper (WASP)             \\
54,392.46445  & 54,392.46520  & $\pm$0.00040  &  29086.0  &  $-$0.00141  &  I   &  This paper (WASP)             \\
54,392.66825  & 54,392.66900  & $\pm$0.00010  &  29086.5  &  $+$0.00034  &  II  &  This paper (WASP)             \\
54,393.67764  & 54,393.67839  & $\pm$0.00021  &  29089.0  &  $-$0.00052  &  I   &  This paper (WASP)             \\
54,394.48549  & 54,394.48624  & $\pm$0.00046  &  29091.0  &  $-$0.00087  &  I   &  This paper (WASP)             \\
54,394.68744  & 54,394.68819  & $\pm$0.00030  &  29091.5  &  $-$0.00096  &  II  &  This paper (WASP)             \\
54,397.51776  & 54,397.51851  & $\pm$0.00090  &  29098.5  &  $+$0.00067  &  II  &  This paper (WASP)             \\
54,398.52796  & 54,398.52871  & $\pm$0.00048  &  29101.0  &  $+$0.00063  &  I   &  This paper (WASP)             \\
54,399.73953  & 54,399.74028  & $\pm$0.00017  &  29104.0  &  $-$0.00010  &  I   &  This paper (AAVSO)            \\
54,402.56812  & 54,402.56887  & $\pm$0.00023  &  29111.0  &  $-$0.00019  &  I   &  This paper (WASP)             \\
54,402.77018  & 54,402.77093  & $\pm$0.00014  &  29111.5  &  $-$0.00018  &  II  &  This paper (AAVSO)            \\
54,402.97262  & 54,402.97337  & $\pm$0.00018  &  29112.0  &  $+$0.00021  &  I   &  This paper (AAVSO)            \\
54,403.98254  & 54,403.98329  & $\pm$0.00022  &  29114.5  &  $-$0.00011  &  II  &  This paper (SOAO)             \\
54,405.39633  & 54,405.39708  & $\pm$0.00026  &  29118.0  &  $-$0.00066  &  I   &  This paper (WASP)             \\
54,405.59869  & 54,405.59944  & $\pm$0.00023  &  29118.5  &  $-$0.00035  &  II  &  This paper (WASP)             \\
54,406.40652  & 54,406.40727  & $\pm$0.00021  &  29120.5  &  $-$0.00072  &  II  &  This paper (WASP)             \\
54,406.60854  & 54,406.60929  & $\pm$0.00033  &  29121.0  &  $-$0.00075  &  I   &  This paper (WASP)             \\
54,407.41721  & 54,407.41796  & $\pm$0.00017  &  29123.0  &  $-$0.00027  &  I   &  This paper (WASP)             \\
54,407.61827  & 54,407.61902  & $\pm$0.00050  &  29123.5  &  $-$0.00126  &  II  &  This paper (WASP)             \\
54,408.22641  & 54,408.22716  & $\pm$0.00040  &  29125.0  &  $+$0.00074  &  I   &  This paper (SOAO)             \\
54,408.42733  & 54,408.42808  & $\pm$0.00020  &  29125.5  &  $-$0.00039  &  II  &  This paper (WASP)             \\
54,410.44898  & 54,410.44973  & $\pm$0.00047  &  29130.5  &  $+$0.00077  &  II  &  This paper (WASP)             \\
54,411.25604  & 54,411.25679  & $\pm$0.00033  &  29132.5  &  $-$0.00036  &  II  &  This paper (SOAO)             \\
54,418.53012  & 54,418.53087  & $\pm$0.00011  &  29150.5  &  $-$0.00003  &  II  &  This paper (WASP)             \\
54,419.33602  & 54,419.33677  & $\pm$0.00056  &  29152.5  &  $-$0.00232  &  II  &  This paper (WASP)             \\
54,419.54028  & 54,419.54103  & $\pm$0.00024  &  29153.0  &  $-$0.00011  &  I   &  This paper (WASP)             \\
54,420.34808  & 54,420.34883  & $\pm$0.00027  &  29155.0  &  $-$0.00051  &  I   &  This paper (WASP)             \\
54,420.55042  & 54,420.55117  & $\pm$0.00063  &  29155.5  &  $-$0.00021  &  II  &  This paper (WASP)             \\
54,424.99560  & 54,424.99635  & $\pm$0.00028  &  29166.5  &  $-$0.00010  &  II  &  This paper (SOAO)             \\
54,426.20819  & 54,426.20894  & $\pm$0.00022  &  29169.5  &  $+$0.00020  &  II  &  This paper (SOAO)             \\
54,429.03623  & 54,429.03698  & $\pm$0.00035  &  29176.5  &  $-$0.00044  &  II  &  This paper (SOAO)             \\
54,429.23965  & 54,429.24040  & $\pm$0.00065  &  29177.0  &  $+$0.00094  &  I   &  This paper (SOAO)             \\
54,429.64266  & 54,429.64341  & $\pm$0.00012  &  29178.0  &  $-$0.00015  &  I   &  This paper (AAVSO)            \\
54,429.84536  & 54,429.84611  & $\pm$0.00012  &  29178.5  &  $+$0.00050  &  II  &  This paper (AAVSO)            \\
54,436.51166  & 54,436.51241  & $\pm$0.00034  &  29195.0  &  $-$0.00079  &  I   &  This paper (WASP)             \\
54,437.52240  & 54,437.52315  & $\pm$0.00025  &  29197.5  &  $-$0.00029  &  II  &  This paper (WASP)             \\
54,438.33104  & 54,438.33179  & $\pm$0.00029  &  29199.5  &  $+$0.00015  &  II  &  This paper (WASP)             \\
54,438.53279  & 54,438.53354  & $\pm$0.00027  &  29200.0  &  $-$0.00014  &  I   &  This paper (WASP)             \\
54,439.34012  & 54,439.34087  & $\pm$0.00021  &  29202.0  &  $-$0.00101  &  I   &  This paper (WASP)             \\
54,439.54268  & 54,439.54343  & $\pm$0.00021  &  29202.5  &  $-$0.00050  &  II  &  This paper (WASP)             \\
54,444.39203  & 54,444.39278  & $\pm$0.00016  &  29214.5  &  $-$0.00030  &  II  &  This paper (WASP)             \\
54,723.8280   & 54,723.82875  & $\pm$0.0001   &  29906.0  &  $-$0.00013  &  I   &  Nelson (2009)                 \\
54,752.31752  & 54,752.31827  & $\pm$0.00007  &  29976.5  &  $-$0.00049  &  II  &  This paper (SOAO)             \\
54,753.12622  & 54,753.12697  & $\pm$0.00025  &  29978.5  &  $-$0.00002  &  II  &  This paper (SOAO)             \\
54,753.32802  & 54,753.32877  & $\pm$0.00006  &  29979.0  &  $-$0.00028  &  I   &  This paper (SOAO)             \\
54,754.33815  & 54,754.33890  & $\pm$0.00019  &  29981.5  &  $-$0.00043  &  II  &  This paper (SOAO)             \\
54,756.96516  & 54,756.96591  & $\pm$0.00010  &  29988.0  &  $-$0.00016  &  I   &  This paper (SOAO)             \\
54,757.16716  & 54,757.16791  & $\pm$0.00010  &  29988.5  &  $-$0.00022  &  II  &  This paper (SOAO)             \\
54,757.97542  & 54,757.97617  & $\pm$0.00011  &  29990.5  &  $-$0.00018  &  II  &  This paper (SOAO)             \\
54,760.19823  & 54,760.19898  & $\pm$0.00005  &  29996.0  &  $+$0.00000  &  I   &  This paper (SOAO)             \\
54,774.94843  & 54,774.94918  & $\pm$0.00008  &  30032.5  &  $-$0.00001  &  II  &  This paper (SOAO)             \\
54,775.15062  & 54,775.15137  & $\pm$0.00009  &  30033.0  &  $+$0.00013  &  I   &  This paper (SOAO)             \\
54,799.3940   & 54,799.39475  & $\pm$0.0005   &  30093.0  &  $-$0.00355  &  I   &  Parimucha et al. (2009)       \\
54,802.6338   & 54,802.63455  & $\pm$0.0009   &  30101.0  &  $+$0.00329  &  I   &  Diethelm (2009)               \\
54,830.31341  & 54,830.31416  & $\pm$0.0002   &  30169.5  &  $+$0.00058  &  II  &  Brat et al. (2009)            \\
54,870.92769  & 54,870.92846  & $\pm$0.00018  &  30270.0  &  $+$0.00026  &  I   &  This paper (SOAO)             \\
55,042.4743   & 55,042.47507  & $\pm$0.0002   &  30694.5  &  $+$0.00025  &  II  &  Parimucha et al. (2009)       \\
55,065.5078   & 55,065.50857  & $\pm$0.0001   &  30751.5  &  $+$0.00017  &  II  &  Parimucha et al. (2011)       \\
55,090.5616   & 55,090.56237  & $\pm$0.0001   &  30813.5  &  $-$0.00002  &  II  &  Parimucha et al. (2011)       \\
55,092.17844  & 55,092.17921  & $\pm$0.00015  &  30817.5  &  $+$0.00044  &  II  &  This paper (SOAO)             \\
55,093.18875  & 55,093.18952  & $\pm$0.00031  &  30820.0  &  $+$0.00051  &  I   &  This paper (SOAO)             \\
55,102.8861   & 55,102.88687  & $\pm$0.0002   &  30844.0  &  $-$0.00044  &  I   &  Diethelm (2010)               \\
55,157.03600  & 55,157.03677  & $\pm$0.00009  &  30978.0  &  $+$0.00064  &  I   &  This paper (SOAO)             \\
55,219.2667   & 55,219.26747  & $\pm$0.00030  &  31132.0  &  $+$0.00047  &  I   &  Parimucha et al. (2011)       \\
55,477.8946   & 55,477.89537  & $\pm$0.0001   &  31772.0  &  $+$0.00004  &  I   &  Diethelm (2011)               \\
55,478.2984   & 55,478.29917  & $\pm$0.00040  &  31773.0  &  $-$0.00028  &  I   &  Parimucha et al. (2011)       \\
55,530.02738  & 55,530.02815  & $\pm$0.00012  &  31901.0  &  $+$0.00088  &  I   &  This paper (SOAO)             \\
55,827.24772  & 55,827.24849  & $\pm$0.00007  &  32636.5  &  $-$0.00030  &  II  &  This paper (SOAO)             \\
55,830.27916  & 55,830.27993  & $\pm$0.00011  &  32644.0  &  $+$0.00042  &  I   &  This paper (SOAO)             \\
55,852.30164  & 55,852.30241  & $\pm$0.00010  &  32698.5  &  $-$0.00036  &  II  &  This paper (SOAO)             \\
55,853.31265  & 55,853.31342  & $\pm$0.00010  &  32701.0  &  $+$0.00041  &  I   &  This paper (SOAO)             \\
55,889.2774   & 55,889.27817  & $\pm$0.00090  &  32790.0  &  $+$0.00048  &  I   &  H\"ubscher \& Lehmann (2012)  \\
55,893.7235   & 55,893.72427  & $\pm$0.00040  &  32801.0  &  $+$0.00150  &  I   &  Diethelm (2012)               \\
56,196.19849  & 56,196.19927  & $\pm$0.00016  &  33549.5  &  $-$0.00004  &  II  &  This paper (SOAO)             \\
56,218.22359  & 56,218.22437  & $\pm$0.00009  &  33604.0  &  $+$0.00009  &  I   &  This paper (SOAO)             \\
56,220.04129  & 56,220.04207  & $\pm$0.00013  &  33608.5  &  $-$0.00079  &  II  &  This paper (SOAO)             \\
\enddata
\tablenotetext{a}{HJD in the terrestrial time (TT) scale.}
\end{deluxetable}

\begin{deluxetable}{lcccccc}
%\tabletypesize{\scriptsize}
\tablewidth{0pt}
\tablecaption{Parameters for the LTT orbits of EP And.}
\tablehead{
\colhead{Parameter}      & \multicolumn{2}{c}{Two-LTT}                         && \multicolumn{2}{c}{Quadratic {\it plus} Two-LTT}    & \colhead{Unit}     \\ [1.5mm] \cline{2-3} \cline{5-6}\\ [-2.0ex]
\colhead{}               & \colhead{$\tau_{3}$}       & \colhead{$\tau_{4}$}   && \colhead{$\tau_{3}$}       & \colhead{$\tau_{4}$}   &                                                                 
}
\startdata                                                                                                                                                                                              
$T_0$                    &  \multicolumn{2}{c}{2,442,638.53720(26)}            &&  \multicolumn{2}{c}{2,442,638.52506(23)}            &  HJED              \\
$P$                      &  \multicolumn{2}{c}{0.4041087212(88)}               &&  \multicolumn{2}{c}{0.4041087955(80)}               &  d                 \\
$a_{12}\sin i_{3,4}$     &  3.544(47)                 &  0.719(64)             &&  1.960(71)                 &  0.689(57)             &  AU                \\
$\omega$                 &  85.7(2.5)                 &  35.0(5.5)             &&  170.7(1.5)                &  40.8(5.2)             &  deg               \\
$e$                      &  0.267(61)                 &  0.37(11)              &&  0.480(63)                 &  0.32(10)              &                    \\
$n$                      &  0.014975(76)              &  0.53886(51)           &&  0.022090(62)              &  0.53743(45)           &  deg d$^{-1}$      \\
$T$                      &  2,430,043(124)            &  2,444,220(10)         &&  2,438,864(45)             &  2,443,534(9)          &  HJED              \\
$P_{3,4}$                &  65.82(34)                 &  1.8291(17)            &&  44.62(13)                 &  1.8340(15)            &  yr                \\
$K$                      &  0.02047(27)               &  0.00395(35)           &&  0.00997(36)               &  0.00386(32)           &  d                 \\
$f(M_{3,4})$             &  0.01027(15)               &  0.1110(99)            &&  0.00378(14)               &  0.0972(80)            &  M$_\odot$         \\
$M_{3,4} \sin i_{3,4}$   &  0.367                     &  0.950                 &&  0.253                     &  0.897                 &  M$_\odot$         \\
$a_{3,4} \sin i_{3,4}$   &  17.406                    &  1.640                 &&  13.92                     &  1.577                 &  AU                \\[1.5mm]
$A$                      &                            &                        &&  \multicolumn{2}{c}{$2.815(27)\times 10^{-11}$}     &  d                 \\
$dP$/$dt$                &                            &                        &&  \multicolumn{2}{c}{$5.088(49)\times 10^{-8}$}      &  d yr$^{-1}$       \\[1.5mm]
$\sigma _{\rm all} ^a$   &  \multicolumn{2}{c}{0.00725}                        &&  \multicolumn{2}{c}{0.00674}                        &  d                 \\
$\sigma _{\rm pc} ^b$    &  \multicolumn{2}{c}{0.00079}                        &&  \multicolumn{2}{c}{0.00079}                        &  d                 \\
$\chi^2 _{\rm red}$      &  \multicolumn{2}{c}{1.102}                          &&  \multicolumn{2}{c}{0.981}                          &                    \\
\enddata
\tablenotetext{a}{rms scatter of all residuals.}
\tablenotetext{b}{rms scatter of the PE and CCD residuals.}
\end{deluxetable}

\begin{deluxetable}{lcccccc}
\small
\tablewidth{400pt}
\tablecaption{Applegate parameters for the magnetic activities of EP And.}
\tablehead{
\colhead{Parameter}       & \multicolumn{2}{c}{Long-term ($\tau_{3}$)}      && \multicolumn{2}{c}{Short-term ($\tau_{4}$)}     & \colhead{Unit}       \\ [1.0mm] \cline{2-3} \cline{5-6} \\[-2.0ex]
                          & \colhead{Primary}      & \colhead{Secondary}    && \colhead{Primary}      & \colhead{Secondary}    &
}
\startdata
$\Delta P$                & \multicolumn{2}{c}{0.1342}                      && \multicolumn{2}{c}{1.2641}                      &  s                   \\
$\Delta P/P$              & \multicolumn{2}{c}{$3.84\times10^{-6}$}         && \multicolumn{2}{c}{$3.62\times10^{-5}$}         &                      \\
$\Delta Q$                & ${1.61\times10^{49}}$  & ${4.18\times10^{49}}$  && ${1.51\times10^{50}}$  & ${3.94\times10^{50}}$  &  g cm$^2$            \\
$\Delta J$                & ${7.49\times10^{46}}$  & ${1.45\times10^{47}}$  && ${7.06\times10^{47}}$  & ${1.37\times10^{48}}$  &  g cm$^{2}$ s$^{-1}$ \\
$I_s$                     & ${2.54\times10^{53}}$  & ${1.52\times10^{54}}$  && ${2.54\times10^{53}}$  & ${1.52\times10^{54}}$  &  g cm$^{2}$          \\
$\Delta \Omega$           & ${2.95\times10^{-7}}$  & ${9.55\times10^{-8}}$  && ${2.78\times10^{-6}}$  & ${9.00\times10^{-7}}$  &  s$^{-1}$            \\
$\Delta \Omega / \Omega$  & ${1.64\times10^{-3}}$  & ${5.31\times10^{-4}}$  && ${1.54\times10^{-2}}$  & ${5.00\times10^{-3}}$  &                      \\
$\Delta E$                & ${4.42\times10^{40}}$  & ${2.77\times10^{40}}$  && ${3.92\times10^{42}}$  & ${2.46\times10^{42}}$  &  erg                 \\
$\Delta L_{rms}$          & ${9.86\times10^{31}}$  & ${6.18\times10^{31}}$  && ${2.13\times10^{35}}$  & ${1.34\times10^{35}}$  &  erg s$^{-1}$        \\
                          & 0.0257                 & 0.0161                 && 55.41                  & 34.75                  &  L$_\odot$           \\
                          & 0.0216                 & 0.0060                 && 46.57                  & 13.02                  &  L$_{1,2}$           \\
                          & 0.0072                 & 0.0045                 && 2.97                   & 2.50                   &  mag                 \\
$B$                       & 6.7                    & 5.0                    && 100.7                  & 75.1                   &  kG
\enddata
\end{deluxetable}

\end{document}